\documentstyle[preprint,aps]{revtex}
{\begin{document}
\title{A direct proof of completeness of squeezed odd-number states}
\author{S. Chaturvedi\thanks{Permanent address: School of Physics, University
of Hyderabad, Hyderabad 500 046 {\bf India}} }
\address{Department of Physics, University of Queensland \\
Brisbane, {\bf Australia}}
\maketitle
\begin{abstract}
A direct proof of the resolution of the identity in the odd sector of the Fock 
space in terms of squeezed number states $D(\xi)|2m+1>; D(\xi) = 
\exp(({\xi}a^{\dagger2}-{\xi}^*{a^2})/2)$ is given. The proof entails 
evaluation of an integral involving Jacobi polynomials. This is achieved by the
use of Racah identities.
\end{abstract}
\vskip1.5in
\noindent {\bf PACS No. : 42.50 Dv 03.65 Fd}
\vskip0.5in
\newpage
The squeezed number states [1]
\begin{equation}
|\zeta;n > \,\,\, =\,\,\, D(\xi)|n> \,\,\,; \,\,\, D(\xi) \,\,\,=\,\,\, \
exp\left(\frac{1}{2}(\xi{a^{\dagger2}} - {\xi}^{*}{a^2}) \right) \,\,\,,
\end{equation}
regarded as a generalization of the squeezed vacuum [2] have been thoroughly investigated in the quantum optics  literature for their non classical properties. 
From a group theoretic point of view [3] these may be thought of as generalized coherent states associated with the SU(1,1) group and it can be shown that the squeezed number states $|\zeta;2n+1> \,\,\, =\,\,\, D(\xi)|2n+1>$ based on the odd Fock 
states furnish a resolution of the identity as follows
\begin{equation}
X \,\,\, =\,\,\, \frac{1}{2\pi}\int \frac{d^2{\zeta}}{(1-|{\zeta}^2|)^2} |\zeta;2n+1><\zeta;2n+1| \,\,\, =\,\,\, I_{odd} \,\,\,,
\end{equation}
where $\zeta$ and $\xi$ are related to each other in the following manner 
\begin{equation}
\xi \,\,\, = \,\,\,\, |{\xi}| e^{-i\phi} \,\,\, ; \,\,\, \zeta \,\,\, = \,\,\, 
\tanh|{\xi}|  e^{-i\phi} \,\,\,.
\label{eq:3}
\end{equation}
The operator $I_{odd}$ in $(2)$ denotes the unit operator in the odd sector of the Fock space
\begin{equation}
I_{odd}\,\,\, = \,\,\, \sum_{k=0}^{\infty} |2k+1><2k+1| \,\,\,,
\label{eq:4}
\end{equation} 
and the integration in $(2)$ is over the unit disc centered at the origin in the complex $\zeta$-plane.
The proof of this statement makes use of the Schur's Lemma and  consists in demonstrating that the operator $X$ in $(2)$ commutes $D(\xi)$ for all $\xi$ and hence must be equal to some constant c times the unit operator.( For the squeezed states based on even Fock states the constant c turns out to be infinite and hence one does not have a resolution of the identity similar to that in $(2)$. Another elegant proof of $(2)$ in which only the completeness of $D(\xi)|1>$ is made use of to derive the general result in $(2)$ for any odd Fock state may be found in [4]. In this letter we show that when one attempts a direct proof of $(2)$ one encounters an integral involving Jacobi polynomials and in order to complete the proof one must show that this integral is equal to one. We show that this is indeed so by calculating the integral in question by making use of the Racah identities [5] in the intermediate steps of the calculation.

The operator $X$ in $(2)$, on using the resolution of the identity in terms of the Fock states and the fact that the matrix elements $<2p|D(\xi)|2n+1>$ vanish, may be written as  
\begin{equation}
X \, =\, \frac{1}{2\pi}\sum_{p,q=0}^{\infty}\int \frac{d^2{\zeta}}{(1-|{\zeta}^2|)^2}<2p+1|{\zeta};2n+1><{\zeta};2n+1|2q+1>|2p+1><2q+1| \,\,\,.
\end{equation}
The explicit expressions for $<2m+1|{\zeta};2n+1>$ are as follows.
For $m\geq n$
\begin{eqnarray}
\lefteqn{<2m+1|{\zeta};2n+1> = } \nonumber \\
 & & e^{-i(m-n)\phi}\sqrt{\frac{\Gamma(n+1)\Gamma(m+3/2)}{\Gamma(m+1)\Gamma(n+3/2)}}(|{\zeta|})^{(m-n)}(1-|{\zeta}|^2)^{3/4} P_{n}^{(m-n,1/2)}(1-2|{\zeta}|^2)\,\,\,,
\end{eqnarray}
For $m\leq n$
\begin{eqnarray}
\lefteqn{<2m+1|{\zeta};2n+1> = }  \nonumber \\
 & & e^{-i(n-m)\phi}\sqrt{\frac{\Gamma(m+1)\Gamma(n+3/2)}{\Gamma(n+1)\Gamma(m+3/2)}}(-|{\zeta|})^{(n-m)}(1-|{\zeta}|^2)^{3/4} P_{m}^{(n-m,1/2)}(1-2|{\zeta}|^2)
\,\,\,.
\end{eqnarray}
Here $P_n^{(\alpha,\beta)}(x)$ denote the Jacobi Polynomials [6,7]. The expressions
above can also be written in terms of the Gegenbauer polynomials $C_n^{\alpha}(x)$ or in terms of the associated Legendre polynomials $P_n^{\alpha}(x) $ using the following relations [6,7]
\begin{equation}
P_{n}^{(\alpha,1/2)}(x) = \frac{\Gamma(n+3/2)\Gamma(\alpha +1/2)}{\Gamma(1/2)
\Gamma(\alpha+n+3/2)} \frac{1}{\sqrt{(1+x)/2}} C_{2n+1}^{(\alpha+1/2)}
(\sqrt{(1+x)/2}) \,\,\,,
\end{equation} 
\begin{equation}
P_{n}^{(\alpha,1/2)}(x) =\frac{1}{2^\alpha} \frac{\Gamma(n+3/2)}
{\Gamma(\alpha+n+3/2)} \frac{((1-x)/2)^{-\alpha/2}}{\sqrt{(1+x)/2}}
 P_{2n+\alpha+1}^{\alpha}(\sqrt{(1+x)/2}) \,\,\,.
\end{equation} 
Using the expressions in $(6)$ and $(7)$ in $(5)$  and performing the integration over $\phi$ one obtains
\begin{equation}
X \, =\, \sum_{p=0}^{\infty}\frac{1}{2}\int_{0}^{1} \frac{d|{\zeta}|^2}{(1-|{\zeta}^2|)^2}|<2p+1|{\zeta};2n+1>|^2 \, |2p+1><2p+1| \,\,\,,
\end{equation}
which may be compactly rewritten as
\begin{equation}
X \, =\, \sum_{p=0}^{n}{\cal{I}}_{n,p}|2p+1><2p+1| +\sum_{p=n+1}^{\infty}{\cal{I}}_{p,n}|2p+1><2p+1|\,\,\,,
\end{equation}
where
\begin{equation}
{\cal{I}}_{p,n}=\frac{1}{2}\int_{0}^{1} \frac{d|{\zeta}|^2}{(1-|{\zeta}^2|)^2}|<2p+1|{\zeta};2n+1>|^2 \,\,;\,\, p\geq n \,\,\,.
\end{equation}
Thus to prove the resolution of of the identity in $(2)$ we need to show that ${\cal{I}}_{p,n}$ is equal to one for all integer values of $p$ and $n$ with $p\geq n$. To show this we proceed as follows. Substituting the expression for
$<2p+1|{\zeta};2n+1>$ given by $(6)$ into $(12)$ and defining $x = |{\zeta}|^2 $ we find that ${\cal{I}}_{p,n} $ can be written as
\begin{equation}
{\cal{I}}_{p,n}=\frac{1}{2}\frac{\Gamma(n+1)\Gamma(p+3/2)}{\Gamma(p+1)\Gamma(n+3/2)}\int_{0}^{1}\frac{dx}{\sqrt{(1-x)}} x^{(p-n)}
{\left[ P_{n}^{(p-n,1/2)}(1-2x)\right]}^2 \,\,\,.  
\end{equation}
We now make use of the fact that the Jacobi polynomials $ P_{n}^{(p-n,1/2)}(1-2x) $ can be explicitly written out in two different ways [6,7]
\begin{equation}
P_{n}^{(p-n,1/2)}(1-2x) =\sum_{l=0}^{n}\left(\begin{array}{c} p \\ l\end{array} \right)  \left(\begin{array}{c} n+1/2 \\  n-l  \end{array} \right)
(-1)^{(n-l)} x^{(n-l)} (1-x)^l \,\,\,,  
\end{equation}
\begin{equation}
P_{n}^{(p-n,1/2)}(1-2x) = \frac{p!}{n!\Gamma(p+3/2)}\sum_{m=0}^{n} (-1)^m
\left(\begin{array}{c} n \\ m \end{array} \right)\frac{\Gamma(p+m+3/2)}{(p-n-m)!} x^m \,\,\,. 
\end{equation}
Substituting the expression $(14)$ for one of the Jacobi polynomials in the integral in $(13)$ and the expression $(15)$ for the other and carrying out the integral over $x$ using 
\begin{equation} 
\int_{0}^{1} dx~~ x^{p+m-l}(1-x)^{l-1/2} =  \frac{\Gamma(l+1/2)\Gamma(p+m-l+1)}
{\Gamma(p+m+3/2)} \,\,\,,
\end{equation}
one obtains
\begin{equation}
{\cal{I}}_{p,n}=\frac{1}{2} n! p! \sum_{l=0}^{n}
\frac{(-1)^l \Gamma(l+1/2)}{l!(p-l)!(n-l)!\Gamma(l+3/2)}
\sum_{m=0}^{n}
\frac{(-1)^m (p+m-l)!}{m!(n-m)!(p-n+m)!} \,\,\,.
\end{equation}
Changing $m$ to $n-m$, $(17)$ may be written as
\begin{equation}
{\cal{I}}_{p,n}=\frac{1}{2} n! p! \sum_{l=0}^{n}
\frac{(-1)^l \Gamma(l+1/2)}{l!(p-l)!(n-l)!\Gamma(l+3/2)}
\sum_{m=0}^{n}
\frac{(-1)^m (p+n-l-m)!}{m!(n-m)!(p-m)!} \,\,\,.
\end{equation}
Using the Racah identity
\begin{equation} 
\sum_{m} \frac{(-1)^m (u-m)!}{m!(x-m)!(z-m)!} = 
\frac{(-1)^z (u-z)!(u-x)!(x+z-u)!}{x!z!} \frac{\sin(\pi(x-u))}{\pi} \,\,\,,
\end{equation}
with $u=p+n-l$, $x=p$, and $z=n$ to carry out the sum over $m$ in $(18)$ one gets
\begin{equation}
{\cal{I}}_{p,n}=\frac{1}{2} \sum_{l=0}^{n}
\frac{(-1)^{n+l}}{(l+1/2)} \frac{\sin(\pi(l-n))}{\pi l} =
\frac{1}{2} \sum_{l=0}^{n}
\frac{(-1)^l}{(l+1/2)} \frac{\sin(\pi l)}{\pi l} \,\,\,. 
\end{equation}
The factor ${\sin(\pi l)}/{\pi l} $ ensures that only the $l=0$ term survives yielding 
\begin{equation}
{\cal{I}}_{p,n}= 1 \,\,\,.
\end{equation}
This result, written in terms of Gegenbauer polynomials or in terms of associated Legendre functions, reads as follows
\begin{equation} 
\int_{0}^{1}\frac{dx}{x^2} (1-x^2)^{p-n}{\left[C_{2n+1}^{(p-n+1/2)}(x)\right]}^2
=\frac{p!\Gamma(p+3/2)}{n!\Gamma(n+3/2)}{\left[\frac{\Gamma(1/2)}{\Gamma(p-n+1/2)}\right]}^2 \,\,\,,
\end{equation}
\begin{equation} 
\int_{0}^{1}\frac{dx}{x^2}{\left[P_{p+n+1}^{p-n}\right]}^2 = \frac{(2p+1)!}{(2n+1)!} \,\,\,.
\end{equation}
Here $p\geq n$. The integrals $(13)$, $(22)$ and $(23)$, to
the best of our knowledge, are not listed in any of the standard tables of integrals.
\vskip0.5cm
\noindent{\bf Acknowledgements:} I wish to thank Dr B. Bambah for bringing the 
Racah identities to my attention. I am grateful to the University of 
Queensland for the award of the University of Queensland Travel Grant and to 
the Physics department for hospitality. 
\newpage
\noindent{\bf References}
\begin{enumerate}
\item F.A.M. de Oliveira, M.S.Kim, P.L.Knight and V.Bu\u{z}ek, Phys.Rev.{\bf A41}, 2645 (1990).
\item H.P.Yuen, Phys.Rev.{\bf A13}, 2226 (1976).
\item A.Perelemov, {\it {Generalized coherent states and their applications}}, (Springer, Berlin, 1986). 
\item G.S.Agarwal and S.Chaturvedi, submitted to J.Phys {\bf A}.
\item G. Racah, Phys.Rev.{\bf 61}, 186 (1942) \,\,;\,\, ibid {\bf 62},432 (1942).
\item M.A.Abramowitz and I.A.Stegun {\it {Handbook of mathematical functions}}
, (Dover, N.Y., 1970)
\item I.S.Gradsteyn and I.M.Ryzhik {\it {Tables of integrals, series and products}}, (Academic Press, N.Y., 1965).
\end {enumerate}
\end{document}